\documentclass[twocolumn]{aastex701} 
\usepackage{graphicx}
\usepackage{amsmath}

\received{July 10, 2025} 
\revised{September 05, 2025} 
\accepted{September 11,  2025} 

\newcommand{\p}[1]{{\color{magenta}{#1}}}

\shorttitle{An intense geomagnetic storm originated by stealth Coronal Mass Ejection}
\shortauthors{Vemareddy and Selva Bharathi}

\begin{document}
\title{An intense geomagnetic storm originated from stealth Coronal Mass Ejection: remote and $in situ$ observations by near radially aligned spacecraft}

\author[orcid=0000-0003-4433-8823,sname='P']{P. Vemareddy}
\affiliation{Indian Institute of Astrophysics, II Block, Koramangala, Bengaluru-560 034, India}
\email[show]{vemareddy@iiap.res.in} 
\author[orcid=0009-0006-7652-9754,sname='K']{K. Selva Bharathi}
\affiliation{Indian Institute of Science Education and Research, Tirupati-517 507, India}
\affiliation{Indian Institute of Astrophysics, II Block, Koramangala, Bengaluru-560 034, India}
\email[show]{selvabharathik@students.iisertirupati.ac.in} 

\begin{abstract}
We investigate the solar origin and heliospheric evolution of an intense geomagnetic storm that occurred on March 23–24, 2023. Despite multiple candidate CMEs observed between March 19–21, a weak CME detected on March 19 at 18:00 UT was identified as the cause, originating from the eruption of a longitudinal-filament channel near center of the sun. The channel underwent a smooth transition to eruption phase without detectable low-coronal-signatures. Wide-angle heliospheric imaging revealed asymmetric expansion and acceleration by solar wind drag, achieving an average CME velocity of $\approx$640 km/s. The radial-evolution of the interplanetary coronal mass ejection (ICME) was analyzed by three spacecraft in close radial-alignment. Arrival-times and propagation-speeds were consistent across spacecraft, with 21-hour delay between STEREO-A and WIND attributed to solar-rotation and longitudinal separation. The ICME exhibits magnetic cloud (MC) signatures characterized by right-handed helicity, enhanced density at all three spacecraft. The MC underwent expansion (radial-size increases from 0.08AU at SolO to 0.18AU at STEREO-A), decrease in magnetic field strength with distance; $B_{av}\propto R_H^{-1.97}$ (SolO-STA) and $B_{av}\propto R_H^{-1.53}$ (SolO-WIND). The MC axis is inclined with the ecliptic at $–69^o$ at SolO, $-25^o$ at STA and $-34^o$ at WIND, indicating rotation during heliospheric transit. Importantly, the storm's main phase leads to a peak intensity ($SYM-H=-169$nT) occurring at 24/02:40UT followed by second peak ($SYM-H=-170$nT) at 24/05:20UT due to density enhancement towards MC's tail. The study emphasises the significant geoeffectiveness of weak, stealth CMEs with southward Bz and density enhancements.

\end{abstract}

\section{Introduction}
\label{Intro}

Geomagnetic storms are major disturbances in the Earth's magnetosphere which are measured with the Disturbance Storm Time (Dst) index. Based on the peak Dst index, the storms are classified as moderate (-100 $\le Dst $ $\le$ -50 nT), intense (-250 $\leq$ Dst $< -100 nT$), and superstorms (Dst $\leq -250\,\text{nT}$) \citep{Gonzalez1994_GMS}. These storms are known to be driven by solar wind structures such as interplanetary coronal mass ejections (ICMEs), and/or co-rotating interaction regions (CIRs) that posses a strong southward magnetic field component \citep{Gosling1991, gonzalez1999interplanetary}. During the storm, the IMF field reconnects with Earth's magnetic field, permitting the solar energetic particles to penetrate the Earth's atmosphere. Severe storms can lead to considerable harm, including widespread power failures from grid disruptions, damage to satellites and instruments from surface charging, interference with satellite navigation and radio communications, and major health threats to astronauts \citep{oughton2017quantifying}.  Thus, from the perspective of space weather and space-dependent technology, studying the solar origins and the impact of the geomagnetic storms has been important scientific research.

According to several reports, the most severe storms are triggered by powerful CMEs that come from the vicinity of the solar disk center \citep{Howard1982_Obs_cor_trans,gopalswamy2005}. To understand the solar origins of a geomagnetic storm, one must investigate i) the in situ magnetic field and plasma measurements for ICME signatures, ii) the near-sun imaging observations of its solar counterpart, namely, the CME from the previous 3-4 days, and iii) the CME's source region on the solar disk. Using remote sensing data from SoHO \citep{brueckner1995}, STEREO \citep{kaiser2008}, and Solar Dynamics Observatory (SDO; \citealt{pesnell2012}), numerous studies have indicated that white-light observations of CMEs are associated with solar disk features like filament or prominence structures (e.g., \citealt{Webb1987_ActCMEProm, Gopalswamy2003_PromEru, Vemareddy2012_FilErup, Vemareddy2017_PromEru, Vemareddy2024_PromEru}), X-ray sigmoids (e.g., \citealt{canfield1999, Vasantharaju2019_FormEruSig}), and extreme ultraviolet (EUV) hot channels (e.g., \citealt{ChengX2013, vemareddy2014_IniErup_11719, Vemareddy2022_hotchan}). The appearance of these features is typically accompanied by a solar flare classified as GOES C, M, or X. Based on these observations, the solar sources of a geomagnetic storm are usually recognized by low coronal signatures (LCS) like filament/prominence eruptions, coronal dimming, post-eruption arcades, flare ribbons etc.

The study by \citet{ZhangJie2007_GeostCME} found that 11\% of the 88 intense geomagnetic storms examined showed no eruptive signatures on the solar disk. This indicates that some CMEs can generate significant geoeffects without any obvious large-scale structures (LCS) associated with them. Furthermore, a study by \citet{Richardson2010_ICME} on  ICMEs revealed that there are solar counterparts for several ICMEs that do not have identifiable solar sources. The CMEs that are characterized by weak or absent LCS are referred to as ``stealth" CMEs. It is generally understood that stealth CMEs are not linked to filament or sigmoid eruptions and primarily originate from quiet regions of the Sun, which feature relatively complex yet weak magnetic field distributions \citep{pevtsov2012coronal}. These CMEs typically produce quiet eruptions that have speeds at the lower end of the spectrum (less than 300 km/s). However, high-cadence, high-resolution extreme ultraviolet (EUV) observations, complemented by advanced image-processing techniques, can detect faint LCS from these CMEs \citep{alzate2017}. 

The first observation of a stealth coronal mass ejection (CME) was made in June 2008 \citep{Robbrecht2009_CME_noLCS}. In situ parameters recorded by STEREO-B at a distance of 1 AU revealed a classical magnetic flux rope structure. Notably, this CME did not originate from any active region, as traced back to its source. Instead, it emerged from a quiet area of the Sun with no prominent disk counterparts (LCS) and appeared extremely faint in coronagraph observations. All subsequent stealth CMEs observed since this event have exhibited the characteristic flux rope structure \citep{OKane2021_Stealth_CME}. Given the substantial evidence from detailed in situ observations that these CMEs can cause significant geomagnetic disturbances \citep{Nitta2017_CMEs_noLCS, Nitta2021_Problem_GMS}, it is essential to investigate stealth CMEs and their evolution through the heliosphere for accurate geomagnetic storm forecasts. A thorough qualitative observational analysis should be conducted to study their origins, heliospheric propagation, and in situ characteristics.

Stealth CMEs have been associated with driving intense geomagnetic storms, and their low detectability in near-Sun observations can lead to inaccurate predictions. Effective storm forecasting requires knowledge of the CME's speed at 0.1 AU and the precise location of its source region on the Sun. In this article, we examine an intense geomagnetic storm that arose from a stealth CME originating near the solar disk center. The recorded geomagnetic storm occurred on March 23-24, 2023, and was classified as intense, with a peak Dst of -163 nT. Importantly, in addition to remote observations, the Solar Orbiter (SolO) was in close radial alignment with STEREO-A {\bf (STA)} and near-Earth spacecraft. This alignment enabled us to assess the radial evolution of the CME at heliocentric distances of 0.5 AU and 0.966 AU, along with WIND measurements at L1. The first magnetometer observations from SolO recorded a stealth CME at 0.8 AU, alongside measurements from Wind at 1 AU and BepiColombo \citep{OKane2021_Stealth_CME, Davies2021_FirstCME_SoLO}. The recent launch of space missions like SolO, BepiColombo, and the Parker Solar Probe (PSP) has made multipoint observations of ICMEs frequent, improving our understanding of their evolution in the heliosphere \citep{Davies2020_RadialLong, Moestl2022_multipoint_obs, Lugaz2022_CME_ME}. An overview of the in situ observations of the storm is presented in section~\ref{sec_overview}, while section~\ref{sec_solsource} discusses the solar source CME and its source region. Observations of the CME's heliospheric propagation are detailed in section~\ref{sec_helprop}. A detailed analysis of in situ observations from radially aligned spacecraft is given in Section~\ref{sec_insitu}. Conclusions from this study with a brief discussion is furnished in Section~\ref{sec_summ}.

\begin{figure*}[!htb]
\centering
\includegraphics[width=.8\textwidth,clip=]{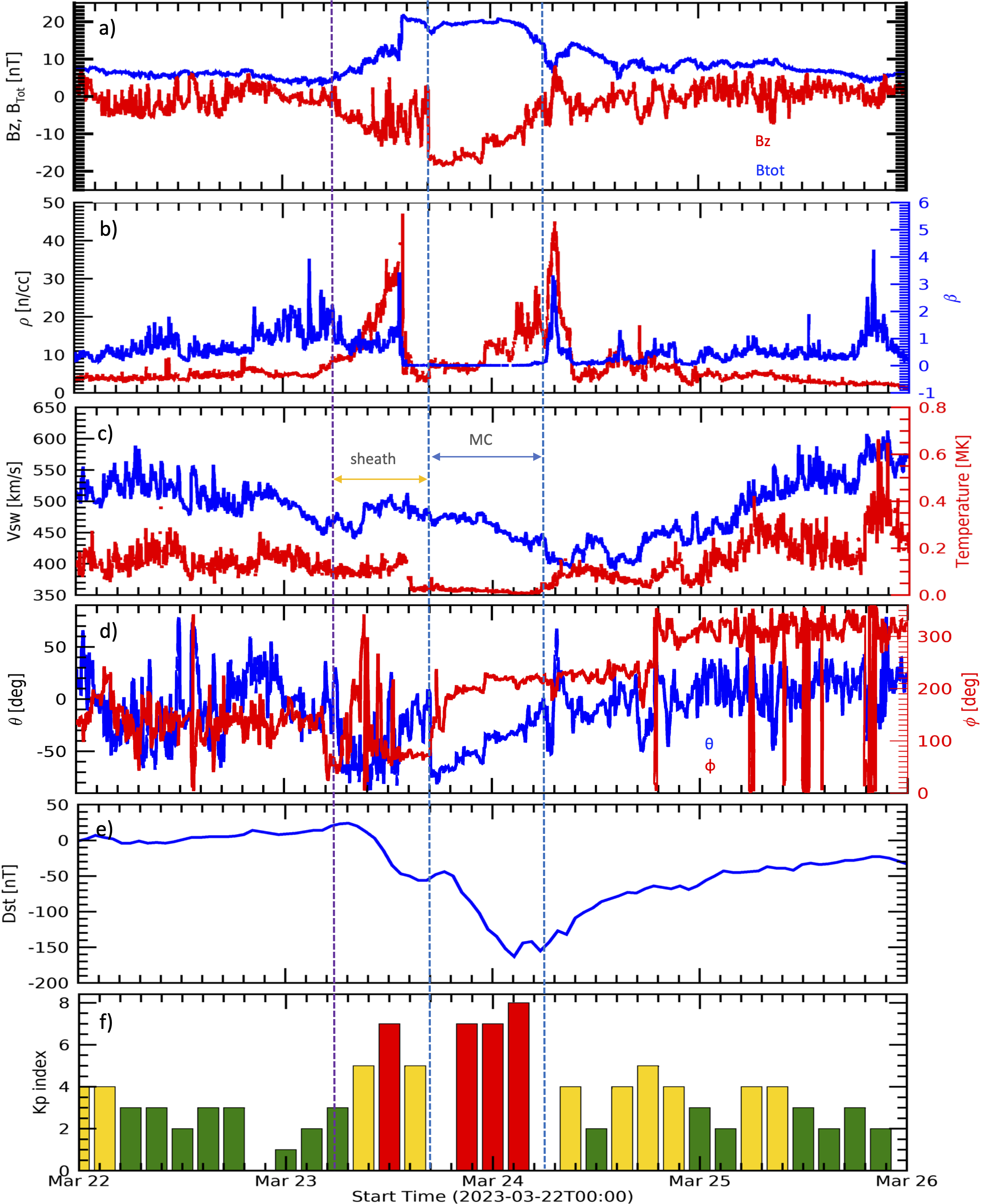}
\caption{An intense geomagnetic storm during March 23-24, 2023 and the in situ observations {\bf (a)} The
Bz and $B_{Tot}$ components of the magnetic field observed by WIND spacecraft. {\bf (b)} The velocity and
the proton density of the ICME. The {\bf purple} dashed vertical line indicates the ICME arrival time (23/07:30 UT), and the blue vertical lines represent the MC interval (23/18:00 - 24/17:30 UT). {\bf (c)} The proton density and proton $\beta$. Note the low density and $\beta$ during MC interval. c) solar wind velocity and temperature. {\bf d)} elevation ($\theta$) and azimuthal ($\phi$) angles of magnetic field vector in GSE frame {\bf (e)} A time-varying Dst index showing the storm commencing at 23/07:30 UT. The main phase of the storm progressed to a peak value of -163 nT on March 24 at 03:00 UT. {\bf (f)} Histogram of Kp index. The green, yellow, and red bars represent the index values 0–3, 3–6, and 6–9, respectively. During the storm's main phase, the Kp index values reached eight, referring to a powerful geomagnetic storm of G4 severity by space-weather classification. }
\label{Fig1}
\end{figure*}

\begin{figure*}[!htb]
\centering
\includegraphics[width=.8\textwidth,clip=]{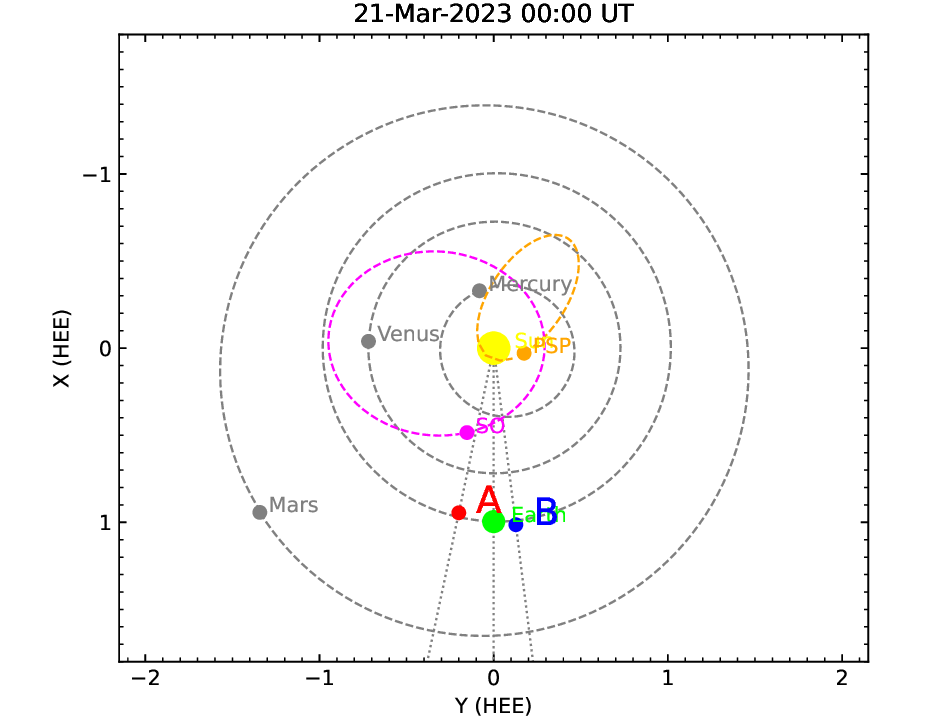}
\caption{The locations of the different spacecraft on 21 March 2023 00:00 UT used in the study. SO (pink) is the Solar Orbiter in its flyby orbit. A (red) and B (blue) refer to STEREO-A and STEREO-B. PSP (orange) near the sun is the Parker Solar Probe in its orbit with the Earth (green) at 1 AU distance from the sun. And the sun (yellow) at [0,0] coordinates.}
\label{sc_loc}
\end{figure*}

\section{Geomagnetic storm during March 23-24, 2023 and near-Earth in situ observations}
\label{sec_overview}
A geomagnetic storm with a Dst value peaking at -163 nT occurred during 23-24 March 2023. Figure~\ref{Fig1} plots the in situ measurements of magnetic field and plasma parameters from WIND, and Dst and Kp indices from NOAA.  While $B_{Tot}$ increasing gradually from background solar wind values, the Bz field is negative and exhibits fluctuations during  07:30-18:00 UT on March 23, which is regarded as the sheath region. Following this, the Bz rotates from negative to positive values while $B_{Tot}$ continuing at 20 nT with little variation until 07:30 UT on March 24. These are the typical characteristics of ICME containing a magnetic cloud. The MC leading and trailing edges have a clear distinction of plasma and magnetic field characteristics, based on which the MC interval is defined from 18:00 UTC on March 23 to 07:30 UT on March 24. The MC structure has low proton density,  proton $\beta$ and temperature compared to its pre and post-passage of spacecraft, as seen in Figure~\ref{Fig1}b-c. During this event, the background solar wind velocity ($V_{sw}$ ) was reportedly higher than 500 km/s, since 14:30 UT on March 21, followed by lesser speed during ICME duration; as a result, there is no shock observed. The MC interval has decreased $V_{sw}$ from 480 km/s to 430 km/s, indicating the expansion of the MC (flux rope), which corresponds to an expansion speed of 25 km/s.

From the magnetic field components, the elevation ( $\theta =sin^{-1}(B_z/B_{Tot}$)) and azimuthal ( $\phi=tan^{-1}(B_y/B_x)$ ) angles are derived and plotted in Figure~\ref{Fig1}d. The $\theta$ and $\phi$ refer to the orientation of the magnetic field vector in the GSE reference frame. The MC leading edge is highly inclined to the ecliptic at $\theta=-80^o$ (south) which then rotates to the ecliptic plane while the azimuth $\phi$ varies from $125^o$ (west) to $240^o$ (east). Therefore, the MC structure is SWN configuration with right-handed (positive) magnetic helicity \citep{Bothmer1998_Struc_MC, Mulligan1998_MagStrMC}. 

Typically, the storm's onset to the main phase corresponds well with the increased southward (Bz) magnetic field; that said, the peak of Bz is co-temporal with the peak of Dst within 2 hours of difference \citep{Vemareddy2024_FilamentEru}. Surprisingly, in this event, the Bz field (magnitude) decreases during the main phase as the Dst progresses to its peak at 24/03:00 UT and the Bz was $Bz=-9.6 nT$, half of its peak value. A key point for this unusually intense storm probably lies with the role of proton density, which is increasing from its lower value towards the MC's trailing edge. MCs with trailing density enhancements were observed in the few reports and are suggested to strengthen the storm \citep{Fenrich1998_MC_DenEnh, Biosi2016_May1998_storm, Gopalswamy2022_WhatIsUnusual}.  

The storm was classified as a G4-intense on the NOAA scale as the Kp index reached a maximum of 8. Ionospheric disturbances were observed over the European sectors as studied by \textcite{Nykiel2024_storm_ionosp} along with disturbances of the GPS/GLONASS signals over the city of Apatity in Russia as studied by \textcite{Belakhovsky2024_GPS_dist}. The storm produced auroras over the USA, extending up to New Mexico and other mid-latitude parts of the world. It was reported that Rocket Lab delayed its launch process by 90 minutes after assessing the effects of the GMS.

Figure~\ref{sc_loc} shows the locations of the sun-observing spacecraft in the Heliocentric Earth Ecliptic (HEE) system in the inner heliosphere on 21 March 2023. SolO was located at 0.5 AU heliocentric distance from the Sun, and is separated by 17.5$^o$ from the Sun-Earth line; WIND was at the L1-point along the Sun-Earth line. STA was 0.966 AU away from the sun and had an angular separation of  11.9$^o$ from the Sun-Earth line. In addition to L1 point, the small angular separation of two spacecraft facilitates the study of the radial and longitudinal variation of magnetic and plasma in the ICME. The radial propagation of that ICME would mean it would encounter SolO, STA and WIND. So, we took advantage of this rare radial lineup of spacecraft and compared the ICME's $in$ $situ$ parameters, flux rope configuration, and its radial evolution at 0.5 AU, 0.966 AU, and at the L1-point.  

\begin{figure*}[!ht]
\centering
\includegraphics[width=0.8\textwidth,clip=]{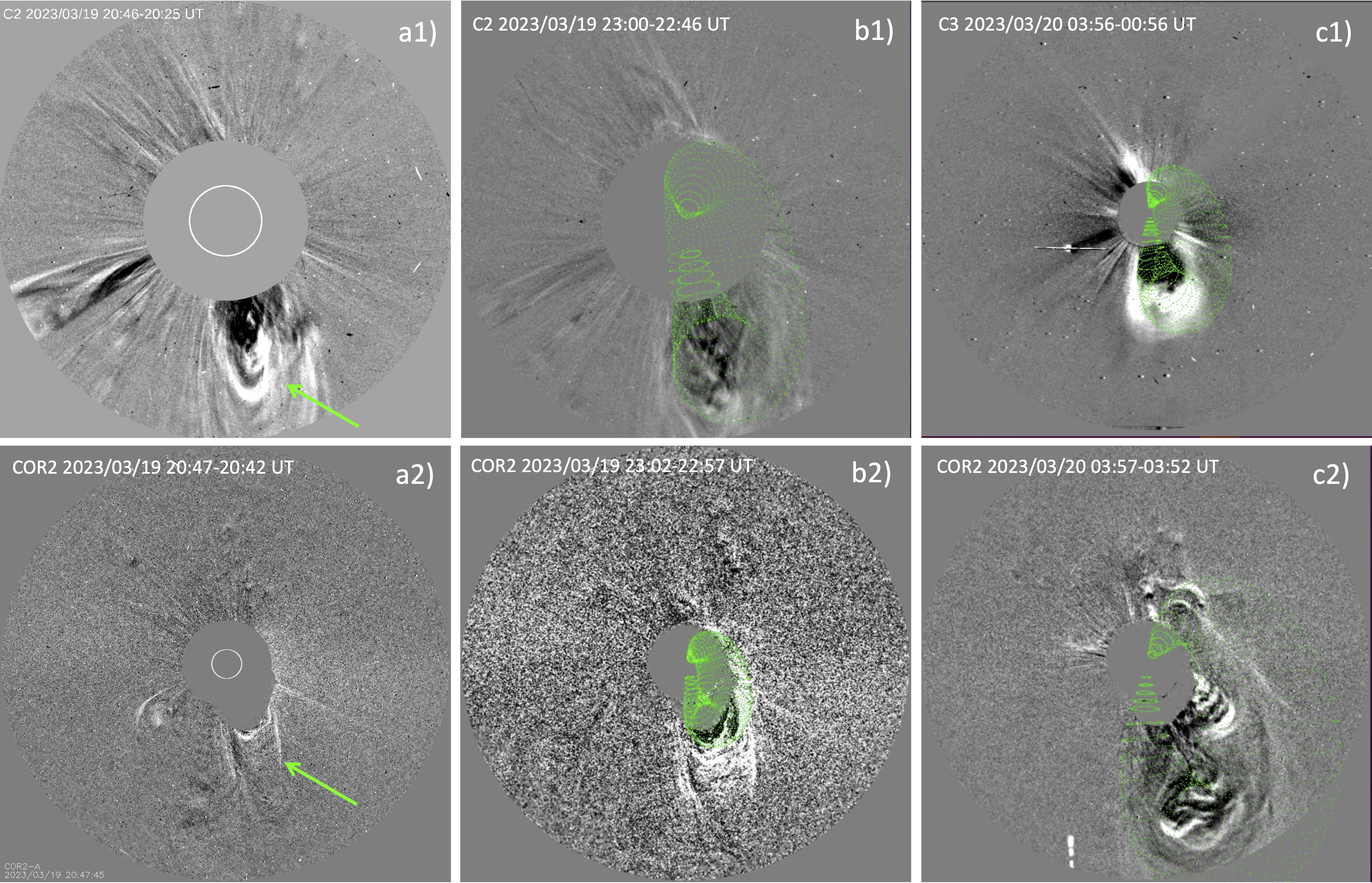}
\caption{Running difference images of coronagraph observations from SOHO/LASCO and STA/COR2. The southern part of the CME from the eruption of the longitudinal filament channel is visible with a bright LE and the core. GCS fit (green wired ) to the CME morphology is overlaid.  }
\label{c2cor2}
\end{figure*} 

\begin{figure*}
\centering
\includegraphics[width=0.92\textwidth,clip=]{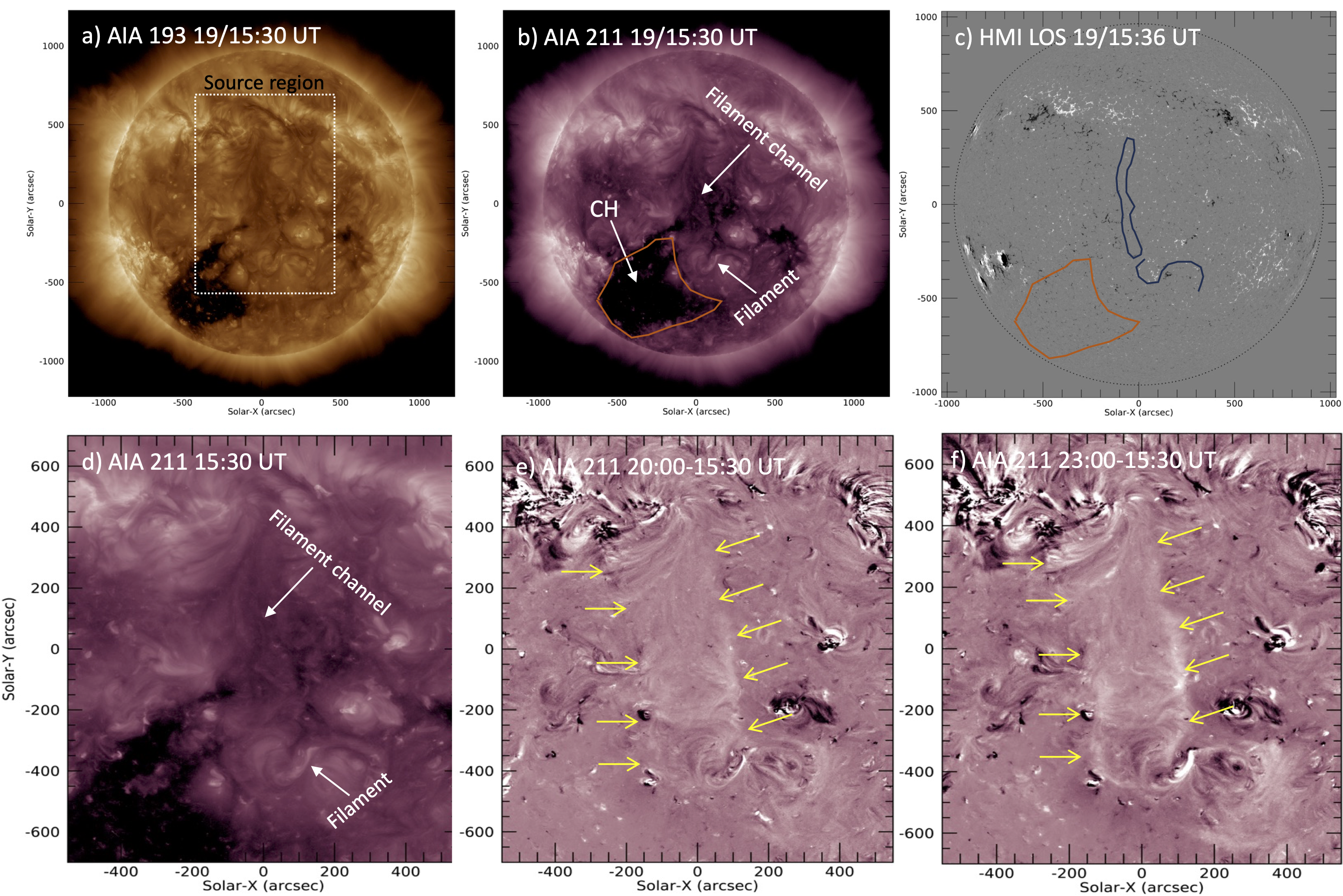}
\caption{Source region observations of the CME (a) Image of the sun at 19/15:30 UT in AIA 193 \AA~channel. Rectangular box encapsulates the source region containing a long trans-equatorial filament channel and a separate S-shaped filament that are being erupted at a later time. b) Sun in AIA 211 \AA~wavelength. The coronal hole, along with the filaments, is referred with the arrows. c) HMI magnetogram showing the magnetic field distribution at the photosphere. To identify the location, the filaments and coronal hole are roughly sketched over the magnetogram. d) Source region in AIA 211 \AA. e-f) Base difference images showing faint bright channels (pointed with yellow arrows) along the extent of filaments being erupted. }
\label{Fig_SourceReg}
\end{figure*}

\section{Solar source of the geomagnetic storm }
\label{sec_solsource}
To identify the solar source of this storm, the remote sensing multiwavelength images captured by SDO, STEREO-A {\bf (STA in short)}, and SOHO are searched for an on-disk eruption and its associated halo white-light CME in the past four days. We have examined the running difference (RD) white-light images available at the CDAW CME catalog\footnote{\url{https://cdaw.gsfc.nasa.gov/CME_list/}} \citep{yashiro2004}. Few potential CMEs occurred on March 19-20, however, their arrival at L1 was ruled out as these CMEs are not halo, and their source region was located near the limb. After a careful examination of running difference images, a faint, slow-moving CME was identified in the LASCO/C2, STA/COR2 observations on March 19 at around 19:00 UT (19/19:00 UT for brevity, and use the same in the rest of this paper). Figure~\ref{c2cor2} shows the CME in the RD images from LASCO/C2 and STA/COR2. In C2 images, the CME first appeared at around 19/18:00 UT in the southern part as a narrow structure propagating southward, with a position angle slightly greater than 200$^o$. As the CME moved out radially, its extent was also seen in the solar north. Mostly the southern LE was traceable during the first two hours of the CME's expansion. Starting from 22:30 UT, a faint structure of this CME appears in the north. In the subsequent LASCO/C2 RD images, as seen in Figure~\ref{c2cor2}, the northern edge of the CME along with the prominent southern part is apparently noticeable, identifying the CME as a halo structure. The CME was not seen to be associated with solar flares and radio bursts. The observed CME morphology was fitted with graduated cylindrical shell (GCS; \citealt{thernisien2009}) model. This delineates the CME a halo structure, although it is seen mostly in the south as a narrow structure. The underlying flux rope at a latitude of -37$^o$ with a tilt angle of -80$^o$, and height of 7.5R$_\odot$ better represents the observed CME morphology (middle panels of) at 23:00 UT. If the latitude is near zero, then the CME might have emerged symmetrically in the south and north parts of the solar disk. This CME flux rope expanded to 20.7 R$_\odot$ by 20/03:57 UT. 

On visually examining the regular EUV images four hours prior to the CME appearance in LASCO/C2, it was found surprisingly that this CME was not linked to any apparent LCS such as flare ribbons, EUV dimming, or post-eruption arcades. Due to weak or unnoticeable low coronal characteristics, this CME falls in the category of stealth-CME directed toward the Earth. 

To determine the source region of the CME, we analyze the EUV imaging observations obtained from SDO/AIA and STA/EUV. AIA captures the full disk of the Sun in 10 different wavelengths with a cadence of 12 seconds \citep{lemen2012}. Figure~\ref{Fig_SourceReg}(a-b) presents full-disk observations of the Sun in AIA 193 and 211 \AA~wavelength channels taken 4 hours prior to the emergence of the CME in LASCO/C2. These images reveal a longitudinal filament channel extending across the equator, along with an S-shaped filament in the southern region. Notably, these features are absent in the AIA 304 \AA~observations, suggesting that the plasma in the filament channel is diffuse. The filament channel measures approximately 1000 arcseconds longitudinally and slightly over 200 arcseconds laterally, appearing as an uneven longitudinal strip that is dimmer than the surrounding quiet regions of the Sun. Additionally, a huge coronal hole exists in its vicinity to the southeast. The magnetic field distribution observed (Panel~\ref{Fig_SourceReg}(c)) indicates that the filament channel formed amidst weak opposite polarities, with positive polarity on the western side and negative polarity on the eastern side of the channel. For the axial field to be directed southward, the helicity must be right-handed, as observed in situ; however, the threads within the channel are not sufficiently pronounced to support this. Minimal emissive activity was detected within the channel between March 18 and 19, and standard EUV observations provided inconclusive results regarding any eruptive signatures. To investigate its connection to the observed CME, we analyzed the EUV images by applying various combinations of differencing cadence. We utilized the base image taken at 15:30 UT, as shown in Figure~\ref{Fig_SourceReg}(d), to subtract from the images captured until the CME observation time in LASCO/C2. Before subtraction, the images are corrected for solar rotation.

Figure~\ref{Fig_SourceReg}(e-f) displays the base difference maps of the source region, capturing the faint brightening along the extent of the channel starting from 18:00 UT and becoming more pronounced after 20:00 UT on March 19. We speculate that the reconnection processes may have created a magnetic structure without filament material that was slowly expelled from the Sun and this structure may have contributed to a CME formation. As a result, the post-eruption arcades form due to reconnection underneath the erupting flux rope (filament). The observed faint brightening along the filament channel over the course of 4 hours suggests that the eruption is slow such that the channel smoothly transitioned to a CME. A slight transverse expansion within the channel was also noted. 
The laterally formed post-eruption arcades and brightenings suggested the presence of a longitudinal flux rope (FR) configuration. These coronal signatures observed between 20:00 UT and 23:00 UT on March 19, indicated a slow and subtle eruption of the longitudinal filament channel resulting in a CME. This narrow longitudinal filament channel resembled the shape and morphology of the CME, as also delineated by the GCS fit, confirming that the LCS observed above corresponded to the CME. Because of the slow eruption, the CME has a low linear speed of around 300 km s$^{-1}$, and underwent slow acceleration during its early phase. A detailed study by \citet{TengWeilin2024} reveals that this storm on March 23, 2023, was indeed associated with the eruption of a longitudinal (trans-equatorial) filament channel on March 19, exhibiting weak LCS and faint CME emission. In the sections that follow, we will provide a detailed analysis of the CME's heliospheric propagation and its in situ observations at varying radial distances.
\begin{figure*}[!htp]
\centering
\includegraphics[width=.95\textwidth,clip=]{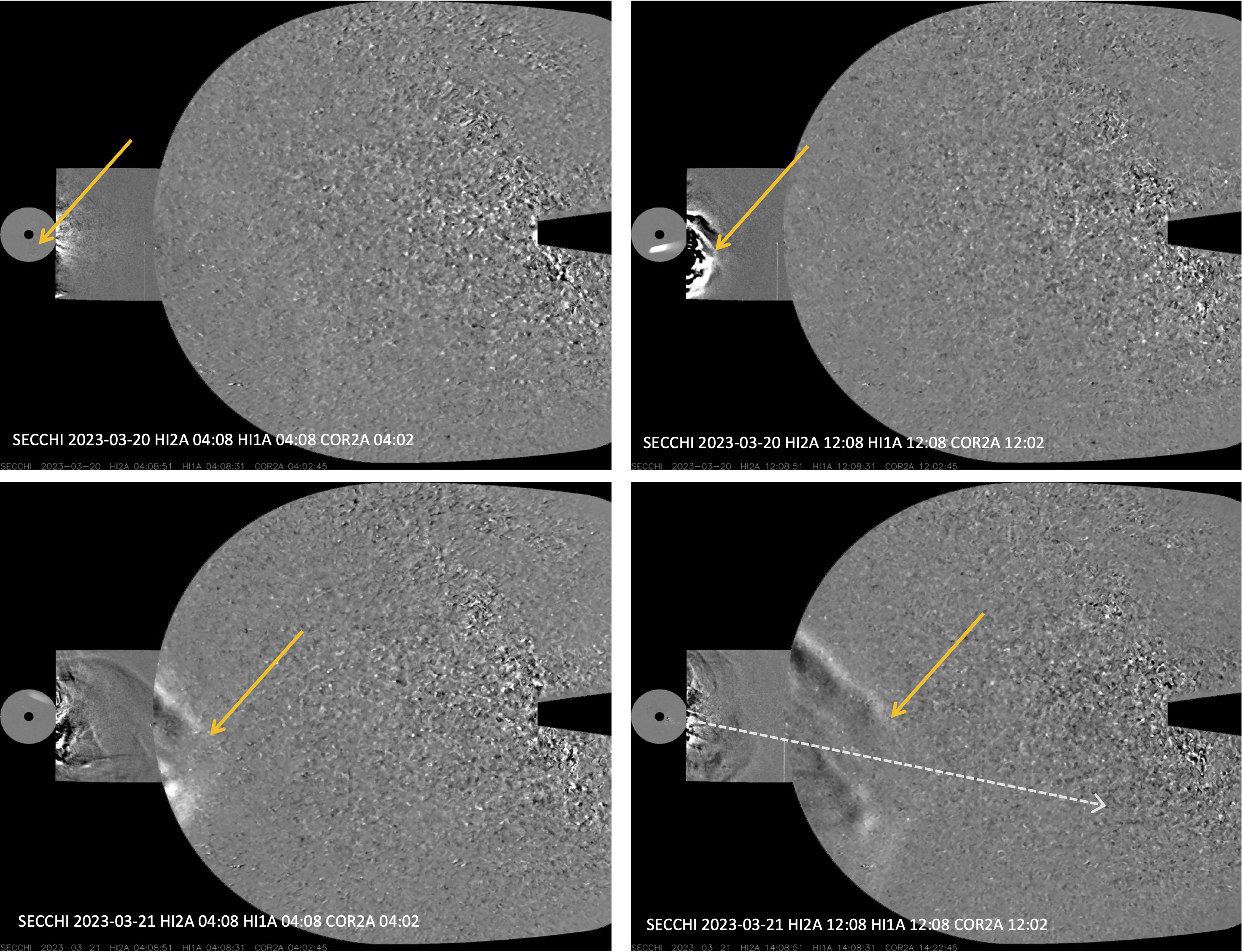}
\caption{CME propagation from the low corona into the heliosphere. Panels show the combined images prepared from running difference images of STA/COR2, STA/HI1, and STA/HI2. The yellow dotted arrows point to the CME's leading edge at different epochs during  March 20-21. White-dashed arrow indicates slit position to construct J-map.   Due to the low separation angle ($12^o$) of STA with the sun-earth line, these images capture the CME propagation with projection effects.}
\label{comb_img}
\end{figure*}
\section{Heliospheric Propagation of the CME}
\label{sec_helprop}
\begin{figure*}[!htb]
\centering
\includegraphics[width=.85\textwidth,clip=]{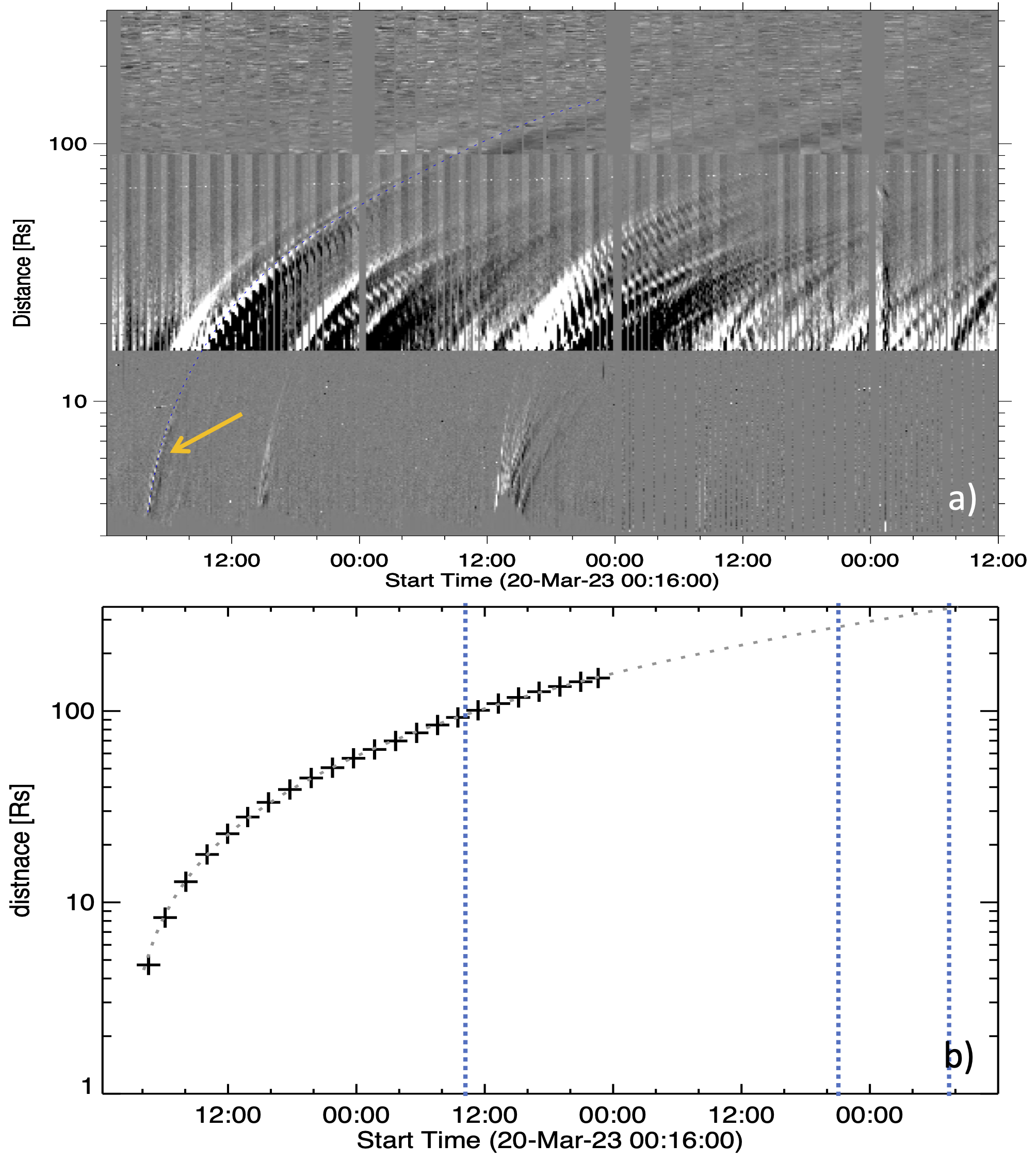}
\caption{(a) The time-elongation map (J-map). The {blue-dotted} curve along the bright streak represent traces of the CME trajectory. (b) The height-time plot derived from the J-map. The ``+'' symbols are the plotted points from the J-map. The dotted curve is the second-order polynomial fit. The vertical dotted lines refer to in situ arrival ICME at SolO (21/09:20UT), STA (22/21:20 UT), and WIND (23/07:30 UT), respectively. }
\label{fig_htplot}
\end{figure*}

The CME's heliospheric propagation was further tracked using the wide-angle observations from the Heliospheric Imagers (HI-1 and HI-2) onboard the STA. The running difference (RD) images from each instrument's observations were combined to create composite images that visualized the CME's lateral expansion as it radially propagates. Figure~\ref{comb_img} shows the composite images at four selected epochs. Visibly, the CME exhibited asymmetrical expansion in the plane of the sky, as we can observe the distorted LE and the southward part advancing faster than the northward LE segment in the HI2 FOV (Figure~\ref{comb_img}(c-d)). Using these composite images, the LE of the CME was tracked where its transit was along the position angle of 250$^o$. The resulting J-map is displayed in Figure~\ref{fig_htplot}.  In the J-map, the CME-LE is identified as a bright streak, which is traced by a blue dotted line. As the CME moves farther away from the Sun, it becomes diffused, and the LE is traceable only upto 130 R$_\odot$. We have fitted the height-time information obtained from this J-map with a second-order polynomial. The ICME arrival time detected by SolO is consistent with the LE propagated to $\approx 100 R_\odot$ in HI. However, the height values corresponding to the ICME arrival at STA and WIND were 240 $R_\odot$ and 320 $R_\odot$, which were inconsistent due to the plane-of-sky projection effects by the halo (I)CME.

From the height-time curve, the CME kinematics are evaluated. In the STA/COR2 FOV, the velocity profile was noted between 00:43 UT and 03:43 UT on 20 March. Between 8.15 $R_\odot$ and 13 $R_\odot$, the velocity of the CME increased from 235.21 $\rightarrow$ 276 $\rightarrow$ 450 km $s^{-1}$. The average velocity was 273 km $s^{-1}$ and its acceleration was 2 m s$^{-2}$. For an extremely slow-moving CME, its acceleration was less and was consistent with the values (-50 to 50 m s$^{-2}$) in the study of \citet{Gopalswamy2011_CME_helio_conseq}. In the HI1 FOV, at 10:00 UT (18 $R_\odot$), 19:46 UT (44.6 $R_\odot$), and 23:04 UT (54.5 $R_s$) on 20 March, the velocities increased from 513.5 $\rightarrow$ 552 $\rightarrow$ 625.5 km s$^{-1}$. The average velocity was 583.3 km s$^{-1}$ with an average acceleration of 3.2 m s$^{-2}$ where both profiles slightly increased. Finally, in the STA/HI2 FOV, between 08:51 UT (90 $R_\odot$) and 15:09 UT (118 $R_\odot$) on 21 March, the velocities further increased to 776 km s$^{-1}$ to 866 km $s^{-1}$. But the speed of the ICME decreased to 786 km $s^{-1}$ to 773.3 km s$^{-1}$ after 18:22 UT (post 131.5 $R_\odot$). The above velocity estimates are after fitting to the height-time curve and have an uncertainty of upto 5 km/s as there could be uncertainty in tracing LE. From these kinematic properties, it is evident that the slow CME from the sun was carried away by the solar wind \citep{cargill2004} with an average velocity of 640 $km/s$ and an average acceleration of 1.84 $ms^{-2}$ in the heliosphere. 

\begin{figure*}[!htb]
\centering
\includegraphics[width=0.98\textwidth,clip=]{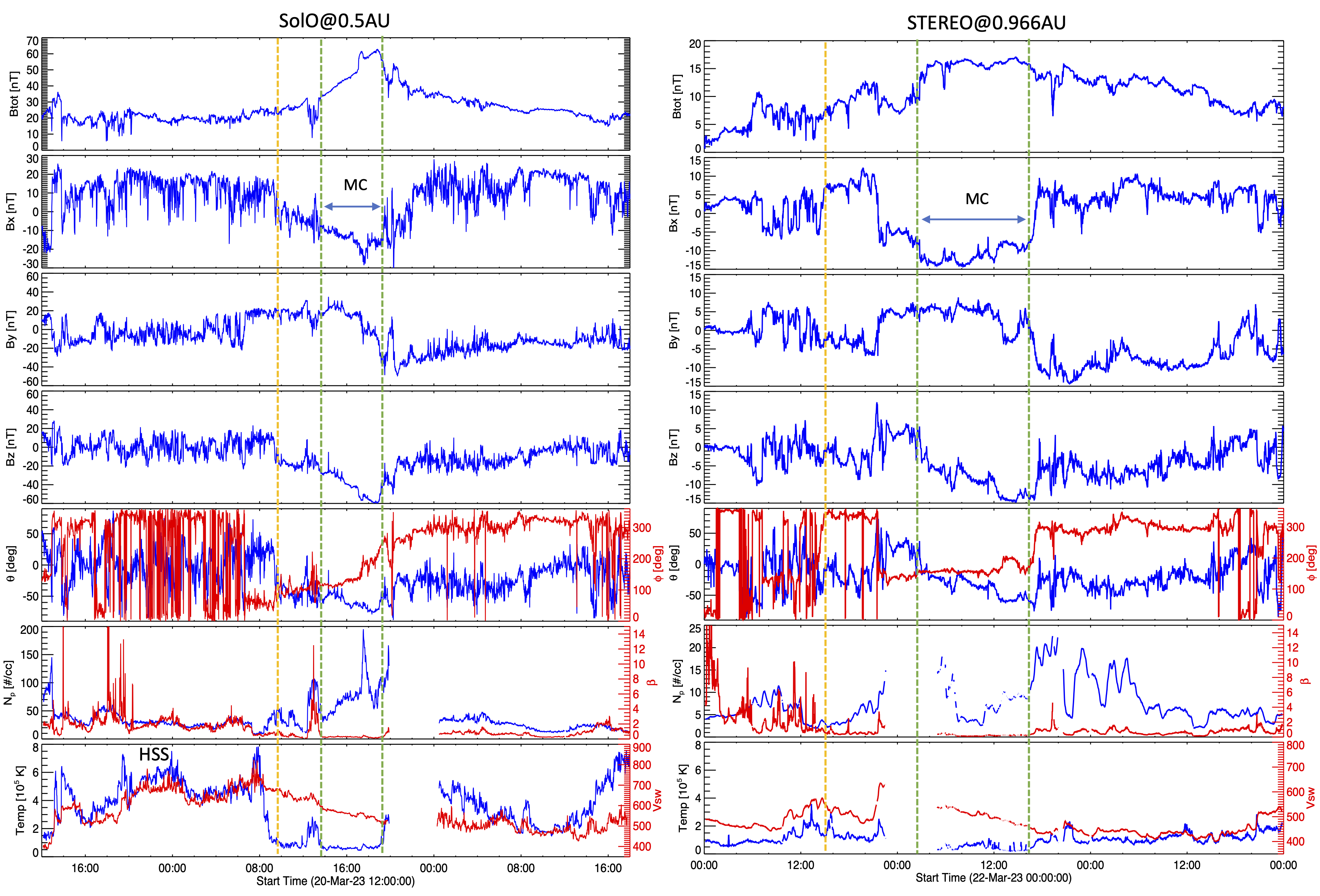}
\caption{Magnetic and plasma measurements at the SolO situated at a radial distance of 0.5 AU and the STA located at 0.966 AU from the Sun. The panels show, from top to bottom, the total magnetic field; the X, Y, Z,  components of the magnetic field in GSE coordinates; the longitude and latitude of the magnetic field angle; the proton density, and $\beta$; and the temperature and velocity. The orange vertical dashed line refers to ICME arrival, and the green vertical lines refer to the leading and trailing edges of the MC. Data gaps are present with velocity and density measurements in both spacecraft.}
\label{fig_sol_sta_comp}
\end{figure*}
\section{In situ observations from SolO, STEREO, and WIND}
\label{sec_insitu}
The $in$ $situ$ magnetic field and plasma parameters are measured by the magnetometer (MAG; \citealt{Horbury2020_SolO_MAG}) instrument aboard SolO, the IMPACT \citep{Luhmann2008_IMPACT}, and PLASTIC \citep{Galvin2008_PLASTIC_instr} instruments on STA, and the MFI \citep{Lepping1995_MAG} and SWA \citep{Ogilvie1995_SWE} instruments onboard WIND spacecraft. For ease of understanding, the magnetic field measurements provided in radial tangential and normal (RTN) coordinate system are approximated to the Geocentric Solar Ecliptic (GSE) coordinate system via $B_x\rightarrow-B_R$, $B_y\rightarrow -B_T$, $B_z\rightarrow B_N$. Figure~\ref{fig_sol_sta_comp} plots the magnetic and plasma measurements by SolO at 0.5 AU heliocentric distance in comparison with STA at 0.96 AU heliocentric distance. 

A high-speed solar wind at an average speed of 680 km/s passes the SolO from 20:00 UT on March 20, followed by the ICME structure. The same high-speed wind, presumed to have originated from the large coronal hole (See Figure~\ref{Fig_SourceReg}), was later observed in WIND from the middle of 21st March. The ICME appears to be encountered at 21/09:20 UT in SolO, with an increased solar wind speed and temperature. The ICME arrival is not noticed with a clear sheath region, but rather a gradual increase of $B_{tot}$ from the background value of 20 nT and without enhanced proton density. The ICME structure hits the STA at 22/21:20 UT, with an enhanced density, temperature, and velocity. In both of these observations, no signatures of shock presence are noticed, probably due to slow CME carried by faster solar wind. The pre-ICME wind velocity decreases from 650 km/s at 0.5 AU to 570 km/s at 0.96 AU. In the WIND observations at 1 AU, the $V_{sw}$ decrease further to 540 km/s, which may possibly be related to both radial and longitudinal evolution from STA. 

Considering an average wind (CME transit) velocity of 650 km/s near the sun (20 R$_\odot$) at 20/02:00 UT, a transit time of 31 hours is consistent with hitting SolO at 21/09:20 UT. This velocity is also in agreement with the observed one in HI field-of-view. A transit velocity of 570 km/s as observed in STA in situ measurements, is justifiable for the CME to travel from SolO to STA in a duration of 36 hours. From the WIND perspective, a transit time of 77 hours is in reasonable agreement with the observed pre-ICME solar wind velocity of 540 km/s.

From the ICME arrival at the spacecraft, low variation of magnetic field components distinct from the background field can be noticed. The variations of the magnetic field are comparable in each component, which is an indication that the same ICME structure passes through the two spacecraft at 0.5 AU and 0.96 AU. According to the criteria for MC structure, its boundary is defined between 13:30-19:30 UT on March 21 in SolO/MAG observations and 02:30-17:00 UT on March 23 in STA/IMPACT measurements. The proton density is higher than background during these intervals, but the proton $\beta$ is quite small for the MC definition. Consistent with the WIND (See Figure~\ref{Fig1}) measurements, the density enhancement towards the trailing edge of the MC is also recorded at both SolO and STA spacecraft. 

Within the MC flux rope, the speed drops from 620 km/s to 525 km/s at SolO and from 580 km/s to 430 km/s at STA, indicating the expanding MC structure as it moves past the spacecraft. Using the midpoint velocity, the radial width of the MC flux rope is calculated by multiplying the time duration by the midpoint velocity. In SolO, it is determined to be 0.08 AU, which changed by about a factor of two to a size of 0.18 AU at STA. Due to the longitudinal separation, the MC size observed at WIND is 0.15 AU, which is smaller than the size recorded at STA.

The MC's expansion speed ($V_{exp}$) is determined using the velocities at the leading and trailing edges as $V_{\text{exp}} = V_{\text{leading}} - v_{\text{trailing}}$. To achieve this, the velocity during the MC duration is linearly fitted, as depicted in Figure~\ref{fig_vexp}, and the slope is taken as the $v_{exp}$. The computed $V_{exp}$ is shown in the corresponding panel. Significantly, the $V_{exp}$ recorded by SolO is 111.25 km/s, which decreases to approximately 50\% at the STA and WIND. A small discrepancy is noted for STA and WIND measurements of $V_{exp}$ (53.12 km/s and 41.0 km/s), likely arising from their positions accessing different regions of the ICME flux rope. These velocities align with the computed radial width of the MC flux rope as it propagates along the sun-earth line.

\begin{figure*}[!htb]
\centering
\includegraphics[width=0.85\textwidth,clip=]{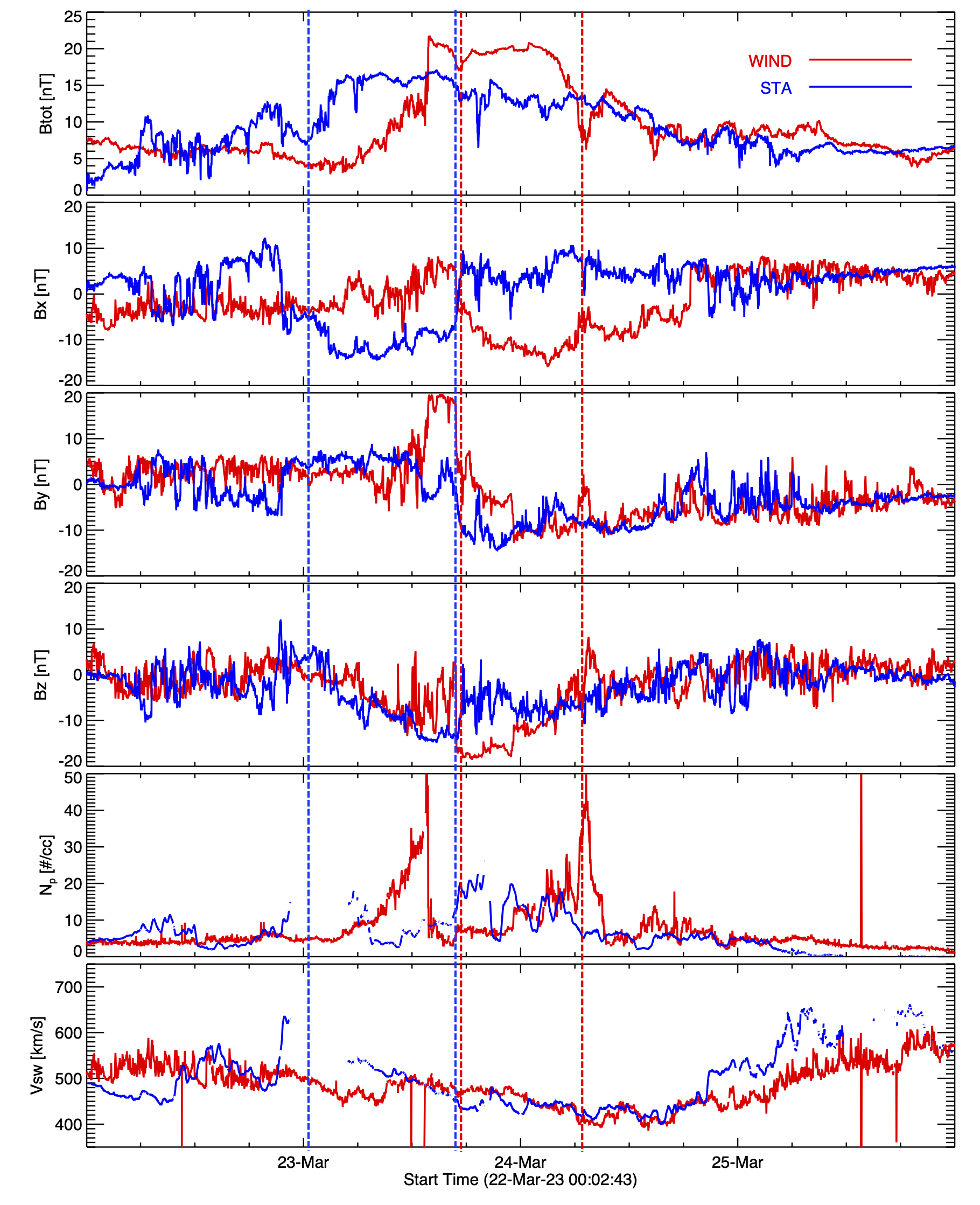}
\caption{ Comparison of solar wind measurements at STA (blue) and WIND (red). The MC interval is marked with vertical dotted lines.  
 }
\label{fig_sta_wind_comp}
\end{figure*}

At a small radial separation of 0.02 AU, Figure~\ref{fig_sta_wind_comp} presents a comparison of in situ measurements from STA with WIND. A significant distinction in these measurements lies in the arrival of the ICME. The ICME meets the STA at 22/21:20 UT and the WIND at 23/07:30 UT. A time difference of 21 hours is clearly evident in the arrival of the MC structure at STA and WIND when we examine the $Bx$ component. We attribute this variation to the co-rotation time between STA and WIND positions \citep{Lugaz2022_CME_ME}. With a co-rotation rate of $14.5^o$ per day, the $12^o$ longitudinal separation between STA and WIND aligns with the recorded time difference in ICME transit. Besides that, there is a significant resemblance between the $Bx$ component and the other magnetic field components observed by both spacecraft. In STA, the $Bz$ component becomes more negative from the leading to trailing edge, while in WIND measurements exhibit an opposite behavior, possibly due to the spacecraft passing away from the MC's center. Unlike STA, the magnetic field components are compressed in front of MC in the WIND due to the high-speed stream. Additionally, the magnetic field strength $B_{tot}$ peaked at approximately $\sim$20.76 nT and $\sim$17.01 nT during the MC interval for the WIND and STA, respectively. These values relate to a power law indicating a decrease in magnetic field with increasing heliocentric distance $B_{peak}\propto R_H^{-2.1}$ from SolO to STA, and a power law of $B_{peak}\propto R_H^{-1.7}$ from SolO to WIND. For the average MC field strength, these relations hold with a slight variation as $B_{av}\propto R_H^{-1.97}$ and $B_{av}\propto R_H^{-1.53}$, respectively.

In the MC structure presented in Figure~\ref{fig_sol_sta_comp}, the $By$ component changes from positive to negative, while the $Bz$ component stays negative in SolO. This led to the elevation angle ($\theta$) of the magnetic vector becoming more inclined toward the south relative to the ecliptic plane, with the azimuthal angle ($\phi$) changing from $90^o$ to $210^o$  in SolO. The noted evolution in $theta$ and $\phi$ indicates that the MC structure possesses a ESW magnetic configuration characterized by right-handed helicity. Unlike SoLO, the $By$ component stays positive, while the $Bz$ component varies from positive to negative at STA, causing the magnetic vector to become more southward as $\phi$ rotates near $140^o$ (east). In this case, the magnetic structure within the MC flux rope is identified as ESW (east-south-west) configuration with right-handed helicity \citep{Bothmer1998_Struc_MC}.  

For the orientation of the MC flux rope, we employed minimum variance analysis (MVA; \citealt{Sonnerup1967_MVA, bothmer1998}), which involves calculating the covariance matrix of the magnetic field components and its eigenvectors. The eigenvector corresponding to the intermediate variance direction gives the orientation of the MC axis \citep{Goldstein1983_MVA}. Although the variations of magnetic fields are similar due to the radial lineup of the spacecraft, the inclination of the MC axis is found to decrease from $\theta=-69^o$ at SolO to $\theta=-25^o$ at STA with respect to the ecliptic plane. For a comparison, this inclination of the MC axis seen by SolO is differed by $20^o$ with respect to the orientation of the filament channel in the solar source region, which is longitudinal with a southward axial field. This decrease in the inclination is a clear signature of the rotation of the MC axis during its radial propagation. Further, even at close radial separation (0.02 AU), a difference in axis inclination $\delta \theta=10^o$ is noticed between WIND and STA. The different characteristics of ICME seen by spacecraft are compiled in Table~\ref{Tab1}. 

\begin{table*}[!ht]
    \centering
        \caption{Characteristics of the ICME observed by three different spacecraft}
    \begin{tabular}{|l|l|l|l|}
    \hline
    Property  &  SolO (0.5 AU)  & STEREO-A (0.966 AU)   & WIND (L1; 0.99 AU) \\
    \hline \hline
    ICME encounter  & 21/09:20 UT    &  22/21:20 UT    &   23/07:30  \\
    MC interval   &  21/13:30-19:30 UT     & 23/02:30-17:00 UT    & 23/18:00-24/07:30 UT      \\
    Velocity at mid of MC    &    560 km/s    &   505 km/s    &   470 km/s   \\
    Expansion  speed         &    111 km/s      &   50 km/s    &   40 km/s   \\
    MC radial size       &  0.08 AU    &  0.18 AU   &  0.15 AU    \\
    Peak $B_{tot}$   &  62.78 nT &  17.01 nT    &   20.76 nT   \\
    mean($B_{tot}$) in MC &   48.2 nT     &  14.2 nT    & 18.0 nT         \\
    MC type       &  ESW   &  ESW     &    SWN       \\
    MC magnetic helicity  &    Right-handed      &  Right-handed       &  Right-handed      \\
    MVA orientation  ($\theta$, $\phi$)    &  $-69^o.5$, $-165^o.9$        &  $-25^o.1$, $-143^o.6$      & $-34^o.5$, $-254^o.2$   \\
    \hline
    \end{tabular}

    \label{Tab1}
\end{table*}

\section{Summary and Conclusions}
\label{sec_summ}
We studied the solar origins of an intense geomagnetic storm that occurred on March 23, 2023. There were multiple candidate CMEs for this storm during 19-21, March, but they were ruled out due to their position of emergence from the limb. A weak CME observed on March 19, at 18:00 UT was identified as the solar source to cause this geomagnetic storm. This weak CME was linked to the eruption of the longitudinal filament channel present at the center of the solar disk from around 18:00 UT onwards, as also studied by \citet{TengWeilin2024}. The filament channel lies along the PIL of distributed weak opposite polarities with a right-handed magnetic twist and southward axial field. Such structures are ideal candidates for producing geoeffective CMEs. Unlike typical eruptions, this filament channel undergoes a smooth transition to the eruption phase, leaving extremely weak LCS. As a result, the CME is not associated with any flare or radio burst. Because of weak or undetectable LCS, the CME is categorized as a stealthy one and is an addition to the stealth CMEs so far identified for their potential geoeffectiveness \citep{Nitta2021_Problem_GMS}.

\begin{figure}[!htb]
\centering
\includegraphics[width=0.49\textwidth,clip=]{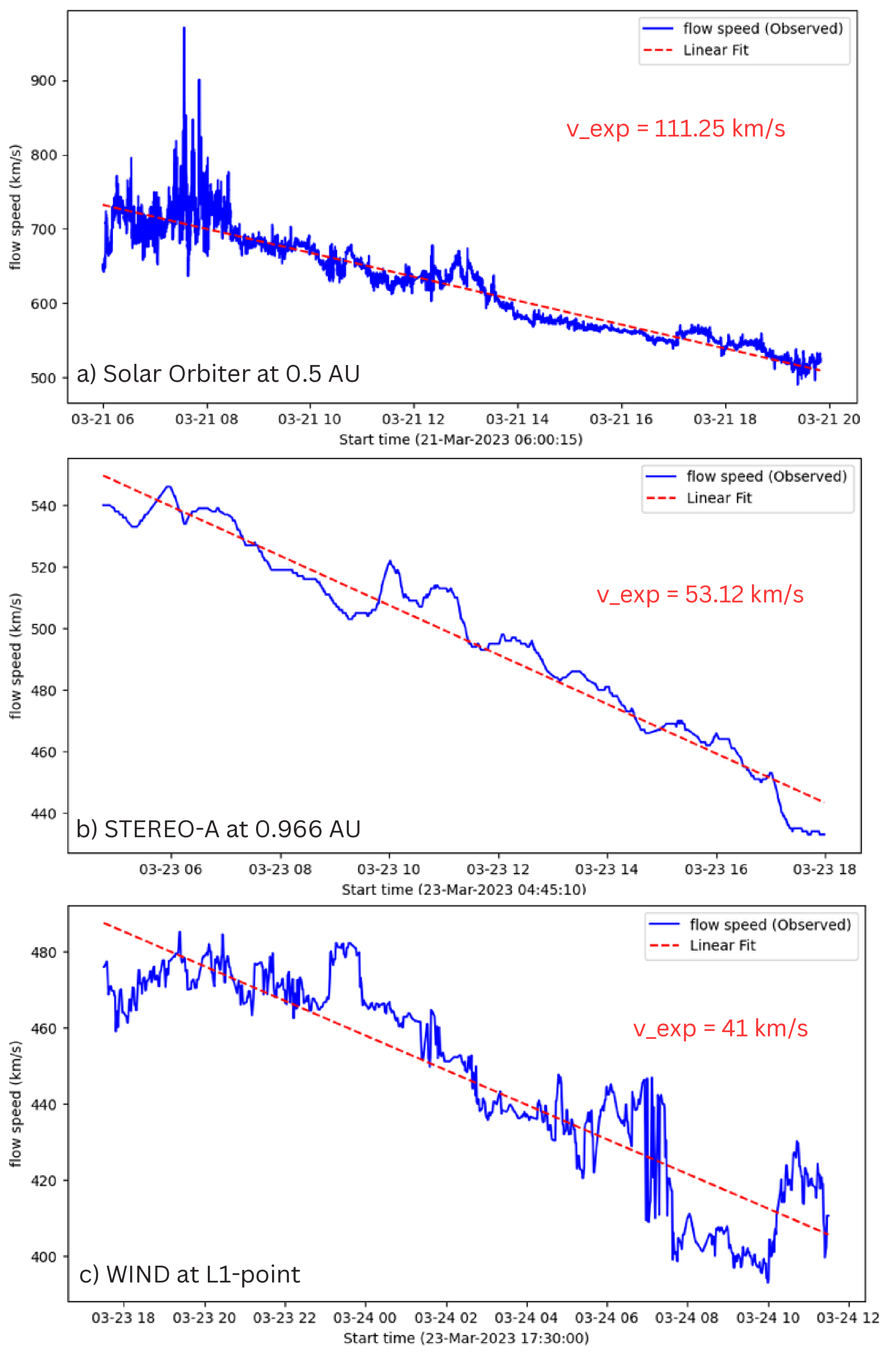}
\caption{The solar wind velocity during the MC interval was observed at three different heliocentric distances by SolO, STA, and WIND spacecraft. The red straight line is a linear fit, and the derived expansion speeds are annotated in each panel. }
\label{fig_vexp}
\end{figure}

\begin{figure*}[!htb]
    \centering
    \includegraphics[width=0.8\textwidth]{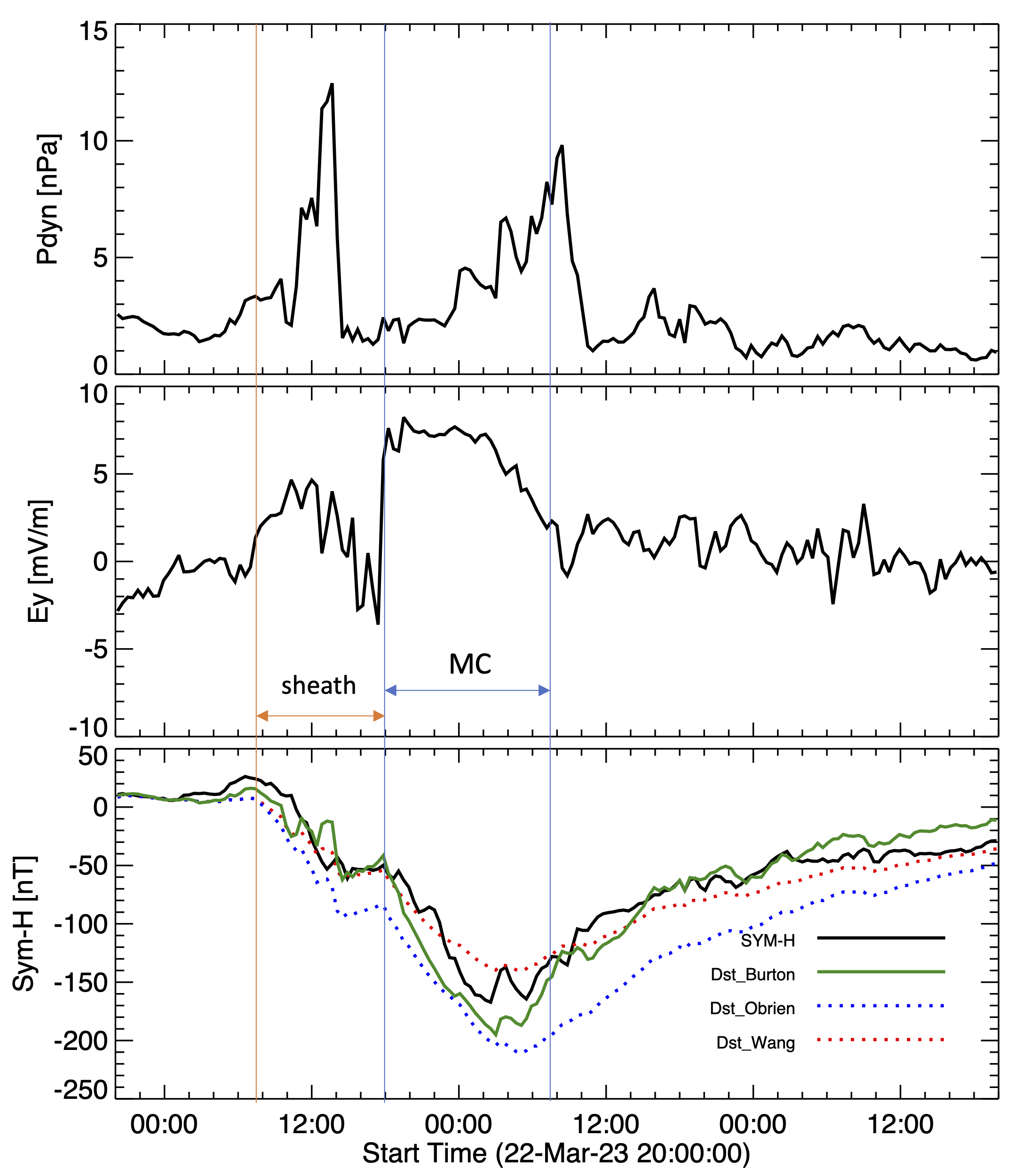}
    \caption{Solar wind dynamic pressure (top), electric field (middle), and observed Dst index (bottom) as a function of time. Estimated Dst index with the models by \citet{Burton1975}, \citet{Obrien2000} and \citet{Wang2003_Influence_DynPres} are overplotted. }
    \label{fig_dst_fit}
\end{figure*}

The heliospheric propagation of the CME was tracked using wide-angle observations from the STA's Heliospheric Imagers. Combined images revealed the CME's asymmetric lateral expansion, with the southern leading edge (LE) advancing faster than the northern segment. The LE was tracked along a position angle of $250^o$, and a J-map traced its motion up to 130 $R_\sun$, showing diffusion with increasing distance. From the height-time data extracted from the J-map, it is inferred that the CME's arrival at SolO matched the propagation seen in HI. Overall, the slow CME was accelerated by the solar wind drag, achieving an average heliospheric velocity of 640 km/s and an acceleration of 1.84 m/s$^2$. 

Further, the radial evolution of the ICME is examined utilizing in situ observations from spacecraft in close radial alignment.  The ICME, preceded by a high-speed solar wind (~680 km/s), was detected without a clear shock or sheath. The ICME's arrival times and propagation speeds were consistent across spacecraft: SolO (0.5 AU) at ~21/09:20 UT, STA (0.96 AU) at 22/21:20 UT, and WIND (0.99 AU) at ~23/07:30 UT. A 21-hour time lag in ICME arrival between STA and WIND is attributed to a $~12^o$ longitudinal separation and solar rotation. The ICME's field component variations are comparable from each spacecraft, with MC signatures suggesting the same ICME structure was encountered. The MC interval observed in all spacecraft shows a decreasing velocity (indicating expansion), a growing radial size (from 0.08 AU at SolO to 0.18 AU at STA), and a reduction in expansion speed—from 111.25 km/s at SolO to about half of that at STA and WIND. A power-law relation describes the observed decrease in magnetic field strength with heliocentric distance, with $B_{peak} \propto R_H^{-2.1}$ (SolO–STA) and $B_{peak} \propto R_H^{-1.7}$ (SolO–WIND).

MC's radial evolution is observed with differences in $B_z$ behavior, which is increasing at STA and decreasing at WIND. The MC magnetic structure is with ESW configuration at SolO and STA, whereas it is with SWN configuration at WIND, all of which implies to a right-handed magnetic helicity consistent with the source region. The MVA indicates that the MC axis is directed at -69 degrees at SolO, closely aligning with the longitudinal filament channel in the source region. However, the inclination of the MC axis reduced to $-25^o$ at STA and $-34^o$ at WIND with respect to the ecliptic, indicating the rotation during MC's propagation in the heliosphere. In summary, the findings emphasize the dynamic expansion, rotation, and structural evolution of the ICME as it propagates through the inner heliosphere.

The Dst index that is observed can be represented by an empirical formula relating the solar wind parameters to the intensity of geomagnetic storms. Commonly used empirical models are \citet{Burton1975} and \citet{Obrien2000} (BM, OM, hereafter) that are based on the assumption that the ring current injection is a linear function of the solar wind's dawn-to-dusk electric field ($E_y$). \citet{Wang2003_Influence_DynPres} (WM hearafter) used an empirical equation providing enhanced Dst estimates, which incorporates the injection term of the ring current based on both the solar wind electric field and the dynamic pressure ($P_{dyn}=n V_{sw}^2$). Figure~\ref{fig_dst_fit}(a-b) shows the $P_{dyn}$ and $E_y$ in as a function of time. The observed (SYM-H) and estimated Dst are plotted in Figure~\ref{fig_dst_fit}(c). During the sheath interval, the $P_{dyn}$ reached a maximum of 13 nPa, and $E_y$ peaked at 5 mV/m, correlating with a SYM-H peak of -67 nT at 23/14:45 UT. Owing to the southward Bz, the electric field $E_y$ maintained a high value of 8 mV/m in the first 6 hours of the MC interval, followed by a steep decline towards the MC tail. In contrast, due to density enhancement towards the MC tail, the $P_{dyn}$ increases from 2 nPa to 8 nPa. The storm's main phase is driven primarily by $E_y$, however, its steep decline later is compensated by the density enhancement from 23/23:00 UT to sustain and further drive the storm until 24/06:00 UT. The main phase of the storm leads to a peak intensity of SYM-H=-169 nT occurring at 24/02:40 UT. The role of density enhancement is clearly indicated by the second peak of SYM-H=-170 nT at 24/05:20 UT. MC structures with enhanced density are suggested to cause intense storms; for example, the 26 August 2018 storm has a steepening Dst profile coincident precisely with the increase of density from 02:00 UT in the MC interval \citep{Gopalswamy2022_WhatIsUnusual}. Empirical models assess the initial, main, and recovery phases of the storm quite effectively, yielding root mean square errors of 18.1, 40.7, and 14.1 nT, respectively, for BM, OM, and WM. The BM overestimates the SYM-H with a peak intensity of -196 nT, while the WM underestimates the storm's peak strength at -152 nT. By incorporating $P_{dyn}$ into the empirical formula, the WM more closely matches the observed Dst, while the BM also achieves a comparable Dst profile by reproducing the two peaks. Our study reports a unique CME that is inconspicuous near the sun propagates to the near-Earth environment with southward Bz and enhanced density together, causing intense geomagnetic activity. 


\

\begin{acknowledgements} SDO is a mission of NASA’s Living With a Star Program; STEREO is the third mission in NASA’s Solar Terrestrial Probes program; and SOHO is a mission of international cooperation between the ESA and NASA. We acknowledge the use of NASA/GSFC's Space Physics Data Facility's OMNIWeb (or CDAWeb or ftp) service. The CME catalog used in this study is generated and maintained by the Center for Solar Physics and Space Weather, the Catholic University of America, in cooperation with the Naval Research Laboratory and NASA. The authors thank the reviewer for a detailed constructive comments and suggestions that helped improve the presentation of the analysis.
\end{acknowledgements}

\end{document}